\documentclass[runningheads]{llncs}
\usepackage[T1]{fontenc}
\usepackage{graphicx,verbatim}
\usepackage{amsmath,amssymb}
\usepackage{multirow}
\usepackage{algorithm}
\usepackage{algpseudocode}
\usepackage{xcolor}
\usepackage[misc]{ifsym}
\usepackage{bbding}

\newcommand{\x}{\mathbf{x}}
\newcommand{\y}{\mathbf{y}}
\newcommand{\z}{\mathbf{z}}
\newcommand{\q}{\mathbf{q}}

\newcommand{\vv}{\mathbf{v}}

\newcommand{\BLC}{\textsc{Bloch}}
\newcommand{\DM}{\textsc{dict-match}}
\begin{document}

\title{Physics informed  guided diffusion for accelerated multi-parametric MRI reconstruction}

\author{
    Perla Mayo\inst{1,}$^{\textrm{\Letter}}$  
    \and Carolin M. Pirkl\inst{2} 
    \and Alin Achim\inst{1}   
    \and Bjoern Menze\inst{3}  
    \and Mohammad Golbabaee\inst{1,}$^{\textrm{\Letter}}$  
}

\institute{University of Bristol, Bristol, United Kingdom \\ \email{\{pm15334, m.golbabaee\}l@bristol.ac.uk} 
\and GE HealthCare, Munich, Germany 
\and University of Zurich, Zurich, Switzerland}

\maketitle              
\begin{abstract}
We introduce MRF-DiPh, a novel physics informed denoising diffusion approach for multiparametric tissue mapping from highly accelerated, transient-state quantitative MRI acquisitions like Magnetic Resonance Fingerprinting (MRF). Our method is derived from a proximal splitting formulation, incorporating a pretrained denoising diffusion model as an effective image prior to regularize the MRF inverse problem. Further, during reconstruction it simultaneously
enforces two key physical constraints: (1) k-space measurement consistency and (2) adherence to the Bloch response model. Numerical experiments on in-vivo brain scans data show that MRF-DiPh outperforms deep learning and compressed sensing MRF baselines, providing more accurate parameter maps while better preserving measurement fidelity and physical model consistency—critical for solving reliably inverse problems in medical imaging. 


\end{abstract}

\section{Introduction}
\label{sec:introduction}
Magnetic Resonance Fingerprinting (MRF) and other transient-state multiparametric mapping techniques~\cite{ma2013mrf,jiang2015fisp,gomez2019qtimrf,cao2022tgas} %
have demonstrated significant advantages over traditional steady-state and/or single-parametric quantitative MRI, enabling faster acquisitions for clinical use. However, additional acceleration through the use of compressed sampling or truncated acquisition sequences, increases image reconstruction artifacts and reduces tissue quantification accuracy. Early MRF reconstruction methods approached this problem with iterative dictionary-matching~\cite{davies2014blip,asslander2018admm,golbabaee2019coverblip}, using sparsity and/or low-rank priors~\cite{mazor2018flor,zhao2018lowrank,lima2019sparsity,golbabaee2021lrtv,cao2022tgas}, 
but struggled with highly undersampled data. Deep learning, particularly convolutional neural networks (CNNs), has since demonstrated superior performance by learning effective image priors from  anatomical datasets~\cite{balsiger2018mrf,fang2019supervisedmrf,fatania2022plug}.

Recently, denoising diffusion models (DDMs) have shown remarkable success in computational imaging~\cite{ho2020ddpm,song2021scorebased,song2021ddim,nichol2021iddpm}. 
In reconstruction, DDMs can use data-driven conditioning to transform artifact-contaminated images into high-quality restored outputs~\cite{dhariwal2021ddpmbeatgans}. 
Recent works~\cite{ajil2021robustcswithdgp,chung2023dps,zhu2023diffpir,liu2023dolce} 
integrate physical acquisition priors, such as enforcing MRI k-space consistency between diffusion steps~\cite{korkmaz2023selfmriunrolleddiff,gungor2023adaptive,peng2022towardsmrdiffusion}, to improve generalization and reliability. Despite promising results in single-parametric qMRI~\cite{bian2024qmri,wang2025qmri} and MRF~\cite{mayo2024ddpmmrf}, current MRF diffusion-based approaches do not explicitly enforce consistency with the physical acquisition model or scanner measurements, which are critical factors affecting reliability.

\textit{Contribution:} To address these limitations, 
we propose MRF-DiPh, a physics-informed diffusion-based MRF reconstruction algorithm. MRF-DiPh uses a pre-trained denoising diffusion model as a spatial image prior to regularize reconstruction, and enforces two essential physical constraints: (1) consistency with k-space measurements and (2) compliance with the Bloch signal-response model. Experiments on retrospectively shortened in vivo brain scans show that MRF-DiPh outperforms existing deep learning and compressed sensing methods, yielding more accurate parameter maps with stronger adherence to physical constraints and scanner measurements.


\section{Preliminaries}
\label{sec:preliminaries}

\textbf{2.1 The MRF Problem:}
MRF reconstruction is a \emph{nonlinear} inverse problem:
\begin{equation}
    \label{eq:forward_operator}
    \y \approx \mathcal{A}(\x)\; 
    \textrm{such that}\;\; \x_v={\boldsymbol\rho}_v  \BLC(\text{T1}_v,\text{T2}_v), \;\forall v : \text{voxels}
\end{equation}
Quantitative T1 and T2 tissue property maps (qmaps) 
have to be estimated for $h \times w$ voxels. $\x \in \mathbb{C}^{s\times h\times w}$ is the time-series of magnetization images (TSMI) that has to be reconstructed. $\BLC(\cdot)$ denotes the nonlinear Bloch response model that voxel-wise encodes qmaps into the time-signals (fingerprints) within the TSMI, scaled by the proton density $\boldsymbol{\rho}$.  
$\mathcal{A}$ is the linear forward acquisition operator that relates TSMI to the undersampled k-space measurements $\y \in \mathbb{C}^{c\times m\times l}$  from $m$ spatial frequency locations across $c$ receiver coil channels, and $l$ time frames. 
$\mathcal{A}$ encompasses coil sensitivities, a nonuniform FFT, and a linear dimensionality reduction ~\cite{mcgivney2014svdmrf,asslander2018admm,golbabaee2021lrtv} which is commonly used for computationally efficient reconstructions by compressing the time dimension of TSMI into adequately smaller $s\ll l$ time frames. 
Additionally, the reconstructions employ an MRF \emph{dictionary} as a discretized approximation of the \BLC\ model, containing a lookup table, $\textsc{LUT}$, of $d$ finely sampled T1-T2 values and their precomputed Bloch responses $D\in \mathbb{C}^{d\times s}$~\cite{mcgivney2014svdmrf}. The Euclidean projections of a TSMI $\x$ onto the Bloch constraints in \eqref{eq:forward_operator} can then be approximated by dictionary matching~\cite{davies2014blip}:
\begin{align}
\label{eq:dm}
& \x_\text{proj} \approx \rho^* D(\text{T1}^* ,\text{T2}^*)\;\; \text{where,}\;\; (\text{T1}^* ,\text{T2}^* ,\rho^* ) \leftarrow \DM(\x) 
\end{align} 
and $\DM(\x) 
= \arg\min_{\rho,(\text{T1,T2})\in \textsc{LUT}}\|\x-\rho D(\text{T1},\text{T2})\|_2$, based on voxel-wise dictionary search using exact or approximate algorithms (see~\cite{cauley2015fastgdm,golbabaee2019coverblip}). 
\newline
\newline
\textbf{2.2 Diffusion Models:}
DDMs generate images from complex distributions $p(\x_0)$  by iteratively refining Gaussian noise samples $\x_T \sim \mathcal{N}(\textbf{0}, \textbf{Id})$ over $t=1,\ldots, T$ diffusion steps~\cite{ho2020ddpm,song2021scorebased,song2021ddim,nichol2021iddpm}. The forward diffusion process corrupts clean images 
$\x_0$ into noisy counterparts $\x_t$ through:
\begin{equation}
\label{eq:sampling_xt}
    \x_t = \sqrt{\bar{\alpha}_t}\x_0 + \sqrt{1-\bar{\alpha}_t}\epsilon \,\,\,\,\text{where,}\,\,\,\, \epsilon \sim \mathcal{N}(\textbf{0}, \textbf{Id}), \;\; t=1,\ldots,T
\end{equation}
where $\{\bar{\alpha}_t\}\in(0,1)$ decreases as $t$ increases and defines a steadily lowering signal-to-noise ratio SNR $:=1/\sigma^2_t=\bar{\alpha}_t/(1-\bar{\alpha}_t)$, until $\x_T$ approximates pure noise. Training a DDM involves learning a \emph{noise estimation network} $\epsilon_\theta$, minimising the loss $\mathbb{E}_{\x_0, t, \epsilon}\left[\| \epsilon - \epsilon_{\theta}(\x_t, t)\|_2^2\right]$. Once trained, 
$\epsilon_\theta$ is used iteratively in a reverse diffusion process to approximate samples from $p(\x_0)$~\cite{song2021ddim}:
\begin{equation}
\label{eq:DDIM}
\x_{t-1} = \sqrt{\bar{\alpha}_{t-1}} \tilde{\x}_{0,t} +\sqrt{1-\bar{\alpha}_{t-1}-\eta^2} \epsilon_\theta(\x_t,t) + \eta\epsilon  
\end{equation}
where $\tilde{\x}_{0,t}:=\frac{1}{\sqrt{\bar{\alpha}_t}}\left(\x_t - \sqrt{1-\bar{\alpha}_t} \epsilon_{\theta}(\x_t, t)\right) $ estimates $\x_0$ by denoising $\x_t$, $\epsilon_\theta(\x_t,t)$ and $\epsilon \sim \mathcal{N}(\textbf{0}, \textbf{Id})$ are the predicted (deterministic) and stochastic noise terms reintroduced to the sample at appropriate scales, and $\eta\in \left[0,\sqrt{1-\bar{\alpha}_{t-1}}\right]$ controls the stochasticity of sampling. 
For inverse problems, reconstruction guidance can be provided by incorporating \emph{conditional} information into training and sampling e.g, a low-quality reconstruction $\x_c=\mathcal{A}^H\y$ from undersampled measurements $\y$ using the adjoint operator. The network is thus parametrized as $\epsilon_\theta(\x_t,t,\x_c)$ to refine denoising adapted to $\x_c$, approximating samples from distribution $p(\x_0|\x_c)$.
\section{MRF-DiPh Algorithm}
\label{sec:methods}
We formulate MRF reconstruction as:
\begin{align}
    \label{eq:general_problem}
    \arg\min_{\x\in \mathcal{B}} f(\x) + \lambda h(\x) 
\end{align}
where $f(\x)= \|\y-\mathcal{A}\x\|_2^2$ enforces k-space consistency, $h(\x)$ regularizes solutions with an image prior, and the regularization parameter $\lambda>0$ balances the two terms. The constraint set $\mathcal{B}:=\{\x \,\,\, \text{s.t.} \,\, x = \rho D(T1,T2)\}$ ensures Bloch model consistency via an MRF dictionary (section 2.1). 
Using Half Quadratic Splitting (HQS), we can solve~\eqref{eq:general_problem} iteratively for $t=T, T-1, \ldots,1$: 
\begin{align}
    &\tilde{\x}_{t} = prox_{\sigma^2 h}(\hat{\x}_t):=\arg\min_{\x} h(\x) + \frac{1}{2\sigma^2} \|\x-\hat{\x}_{t}\|_2^2,  \label{eq:hqs_xtilde} \\
    &\hat{\x}_{t-1} = \arg\min_{\x \in \mathcal{B}} f(\x) + \frac{\mu}{2} \|\x-\tilde{\x}_{t}\|_2^2, \label{eq:hqs_xhat}
\end{align}
where $\sigma^2 := \lambda/\mu$ and $\mu>0$ controls convergence. 
This process decouples image prior enforcement (proximal step \eqref{eq:hqs_xtilde}) from physical acquisition constraints \eqref{eq:hqs_xhat}. 
To incorporate deep learning-based image priors, one could adopt a plug-and-play framework~\cite{zhang2022dpir,ahmad2020pnpmri} 
and replace~\eqref{eq:hqs_xtilde} with a deep denoising model pretrained on images  
to remove additive Gaussian noise of variance $\sigma^2 =1/\text{SNR}$.  Inspired by this idea, we extend HQS with a DDM-based prior: 
\begin{align}
    & \tilde{\x}_{0,t}  =\frac{1}{\sqrt{\bar{\alpha}_t}}\left(\x_t - \sqrt{1-\bar{\alpha}_t} \epsilon_{\theta}(\x_t, t)\right), \label{eq:hqs2:denoise}\\
    &\hat{\x}_{0,t} = \arg\min_{\x\in \mathcal{B}} f(\x) + \frac{\mu_t}{2} \|\x-\tilde{\x}_{0,t}\|_2^2, \label{eq:hqs2_physics}\\
    &\x_{t-1} = \sqrt{\bar{\alpha}_{t-1}} \hat{\x}_{0,t} + \sqrt{1-\bar{\alpha}_{t-1}} \left( \sqrt{1-\xi} \hat{\epsilon}_t + \sqrt{\xi} \epsilon \right)
    \label{eq:hqs2_addnoise}
\end{align}
where $\hat{\epsilon}_t = \frac{\x_t - \sqrt{\bar{\alpha}_t} \hat{\x}_{0,t}}{\sqrt{1-\bar{\alpha}_t}}$ and $\epsilon\sim \mathcal{N}(0,\mathbf{I})$. Eq~\eqref{eq:hqs2:denoise} performs denoising via DDM’s pretrained noise estimator $\epsilon_\theta$, yielding $\tilde{\x}_{0,t}$, an estimate of the clean image $\x_0$ (section 2.2). Eq.~\eqref{eq:hqs2_physics} further enforces k-space and Bloch consistency, producing $\hat{\x}_{0,t}$. Finally, \eqref{eq:hqs2_addnoise} reintroduces noise to the current cleaned image estimate to obtain a diffusion sample $\x_{t-1}$ for the next step. The added noise combines a stochastic $\epsilon$ and a deterministic term $\hat{\epsilon}_t$, balanced by $\xi\in[0,1]$, e.g. $\xi=0$ gives  a fully deterministic sampling.\footnote{The deterministic noise is predicted from the current noisy sample $\x_t$ and the physics-consistent estimate $\hat{\x}_{0,t}$, which differs slightly from the deterministic noise used in~\eqref{eq:DDIM}; replacing $\hat{\x}_{0,t}$ with $\tilde{\x}_{0,t}$ from~\eqref{eq:hqs2:denoise} would recover $\hat{\epsilon}_t=\epsilon_\theta(\x_t,t)$ similar as~\eqref{eq:DDIM}. 
Instead, we update $\hat{\epsilon}_t$ using physics-consistent $\hat{\x}_{0,t}$ from \eqref{eq:hqs2_physics}.}
As discussed in section 2.2, $\epsilon_{\theta}$ estimates noise at various SNR levels  $1/\sigma_t^2$, which increases as sampling progresses, i.e. $t$ decreases. Consequently, $\mu_t := \lambda/\sigma_t^2$ also increases as $t$ decreases.  

The constrained subproblem~\eqref{eq:hqs2_physics} can be solved using the ADMM algorithm~\cite{boyd2011admm}, which introduces an auxiliary variable $\z_t$, the dual variable (Lagrange multipliers) $\vv_t$, and a convergence hyperparameter $\gamma_t>0$. Early experiments showed a single ADMM iteration is sufficient for updating~\eqref{eq:hqs2_physics}. This single-step ADMM update can be written as follows:
\begin{align}
&\hat{\x}_{0,t} = \arg\min_\x f(\x) +  \frac{\mu_t}{2}\|\x-\tilde{\x}_{0,t}\|_2^2 + \frac{\gamma_t}{2} \|\x-\z_t+ \vv_t/\gamma_t\|_2^2, \label{eq:admm_xhat}\\
&\z_{t-1} = \arg\min_{\z\in \mathcal{B}}  \|\z - (\hat{\x}_{0,t}+\vv_t/\gamma_t)\|_2^2 \label{eq:admm_z},\\
&\vv_{t-1} = \vv_t + \gamma_t(\hat{\x}_{0,t} - \z_{t-1}). \label{eq:admm_dualvariable}
\end{align}
where~\eqref{eq:admm_xhat}, equivalent to $prox_{\frac{1}{\mu_t + \gamma_t}f}\left(\frac{\mu_t\tilde{\x}_{0,t}+\gamma_t\z_t-\vv_t}{\mu_t+\gamma_t}\right)$, is a linear least squares term enforcing k-space consistency, which can be efficiently updated by a few conjugate gradient (CG) iterations. Eq~\eqref{eq:admm_z} is a projection onto the Bloch constraints set, approximable via $\DM(\hat{\x}_{0,t}+\vv_t/\gamma_t)$ (see~\eqref{eq:dm}) to give Bloch-consistent TSMI $\z_{t-1}$ and also, the qmaps $\q_{t-1}=(\text{T1}_{t-1},\text{T2}_{t-1},\boldsymbol{\rho}_{t-1})$. 
Finally,~\eqref{eq:admm_dualvariable} updates the ADMM's dual variable $\vv_{t-1}$. 
The ADMM parameter is set as $\gamma_t:=\tau\mu_t$, where $\tau>0$ balances k-space and Bloch consistency constraints. Combining~\eqref{eq:hqs2:denoise}-\eqref{eq:admm_dualvariable}, we obtain MRF-DiPh (Algorithm~\ref{alg:sampling}).
To accelerate reconstruction, we follow~\cite{song2021ddim} and use a sub-sequence $\{t_k\}_{k=1}^K \subseteq [1,\ldots,T]$ of $K\ll T$ time steps, skipping  intermediate steps during diffusion sampling.

\subsection{Implementation Details}
\label{subsubsec:implementation}
Our approach was developed in PyTorch using the guided diffusion toolbox~\cite{github-guided-iddpm}, following the UNet architecture from~\cite{dhariwal2021ddpmbeatgans} at half precision. The UNet consisted of six levels, each with two residual blocks, using channel sizes [128, 128, 256, 256, 512, 512] from highest to lowest feature resolutions, and three attention heads at feature resolutions $28\times28$, $14\times14$ and $7\times7$.

Our base model employed a conditional DDM trained on paired TSMIs $\{\x_c=\mathcal{A}^H\y, \x_0=\x_{\text{ref}}\}$. Low-aliasing reference TSMIs $\x_{\text{ref}}$ were generated from k-space measurements $\y_{\text{ref}}$ acquired through extended MRF scans~\cite{jiang2015fisp}. To simulate faster acquisitions, $\y_{\text{ref}}$ was subsampled by truncation along the time dimension, producing k-space data $\y$, from which highly artifact-contaminated condition images $\x_c$ were obtained. Both $\x_c$ and $\x_{\text{ref}}$ were time-compressed with  
$s=5$~\cite{mcgivney2014svdmrf}. Complex-valued data were processed by concatenating the real and imaginary parts along the channel dimension, and images were range-normalized to [-1,1]. Our framework also supports an unconditional DDM, trained solely on $\{\x_0=\x_{\text{ref}}\}$, without data-driven conditioning/guidance. We explored this approach in Section~\ref{sec:experiments}. Training used $T = 1000$ diffusion steps, $\bar{\alpha}_t=\prod_{t=0}^T (1-\beta_t)$ with linearly spaced $\beta_t$ ($\beta_0 = 0.0001$, $\beta_T = 0.02$), for 100k iterations, using ADAM optimizer with learning rate $10^{-4}$, batch size 32, and data augmentation with random vertical/horizontal spatial flips.
For inference, our base model used experimentally tuned parameters $K = 30, \lambda = 10^{-4}, \tau = 0.01, \xi = 1$, 
CG maximum iterations 5 and initialized by the previous update of line 5 in Algorithm~\ref{alg:sampling}. 

\begin{algorithm}[t]
    \caption{MRF-DiPh
    }
    \label{alg:sampling}
    \begin{algorithmic}[1]

        \Require $f$,  $\epsilon_{\theta}$, $\{t_k\}_{k=1}^K$,$\{\bar{\alpha}_{t_k}\}_{k=1}^{K}, \lambda, \tau, \xi$.
        \State Set $\sigma^2_k:=(1-\bar{\alpha}_{t_k})/\bar{\alpha}_{t_k},\,\mu_k:=\lambda/\sigma_k^2$, $\gamma_k:=\tau \mu_k$
        \State Initialize $\x_K \sim \mathcal{N}(\textbf{0}, \textbf{Id}),\,\mathbf{z}_K=\mathbf{v}_K=0$
        \For{$k=K,\hdots, 1$}
            \State $\tilde{\x}_{0,k} = \frac{1}{\sqrt{\bar{\alpha}_k}}\left(\x_k - \sqrt{1-\bar{\alpha}_k} \epsilon_{\theta}(\x_k, t_k, \x_c)\right)$ \,\textcolor{gray}{//denoised image}
            
            \State $\hat{\x}_{0,k} = prox_{\frac{1}{\mu_t + \gamma_k}f}\left(\frac{\mu_k\tilde{\x}_{0,k}+\gamma_k\z_k-\vv_k}{\mu_k+\gamma_k}\right)$ \hspace{0.45cm} \textcolor{gray}{//k-space consistent image}
            
            \State ($\z_{k-1}, \q_{k-1}) \leftarrow \DM(\hat{\x}_{0,k}+\vv_k/\gamma_k)$\hspace{.2cm}\textcolor{gray}{//Bloch-consistent image \& qmaps}
            
            \State $\vv_{t-1} = \vv_t + \gamma_t(\hat{\x}_{0,t} - \z_{t-1})$ \hspace{2.05cm} \textcolor{gray}{//dual variable update}
            
            \State $\hat{\epsilon}_k = \frac{1}{\sqrt{1-\bar{\alpha}_k}}(\x_k - \sqrt{\bar{\alpha}_k} \z_{k-1})$ \hspace{1.6cm} \textcolor{gray}{//predicted (deterministic) noise}
            
            \State $\epsilon \sim \mathcal{N}(0,I)$ \hspace{3.9cm}\textcolor{gray}{//stochastic noise}
            
            \State $\x_{k-1} = \sqrt{\bar{\alpha}_{k-1}} \z_{k-1} + \sqrt{1-\bar{\alpha}_{k-1}} ( \sqrt{\xi} \epsilon + \sqrt{1-\xi} \hat{\epsilon}_k)$ \textcolor{gray}{//noisy sample for next iter}
        \EndFor ;
        \Return $\x_\text{rec} = \z_0, \q_\text{rec}=\q_0:=\{\text{T1}_0,\text{T1}_0,\boldsymbol{\rho}_0\}$
    \end{algorithmic}
\end{algorithm}

\section{Numerical Experiments}
\label{sec:experiments}

\subsubsection{Setup:} 
The dataset consists of anonymized 2D brain MRF scans from healthy volunteers, obtained with informed consent and acquired using the Steady State Precession (FISP) sequence and flip angles from~\cite{jiang2015fisp} for $l=1000$ time frames, repetition/echo/inversion times of 10/1.908/18 ms, non-Cartesian k-space sampling with variable density spiral readouts, image size $230\times230$, 1mm in-plane resolution and 5mm slice thickness. Scans were performed on a 3T GE MR750w scanner with 8-channel receive-only RF head coils. Models were trained and tested for the reconstruction of $R=5$ fold accelerated scans by retrospective truncation of the FISP sequence length to $l={200}$. The dataset includes 8 subjects with 15 axial slices each, split 75\%-25\% for training and testing. Reference qmaps were reconstructed using LRTV~\cite{golbabaee2021lrtv} from full-length ($l=1000$) scans, from which the reference TSMIs were estimated using $\BLC$ model in~\eqref{eq:forward_operator}.  An MRF dictionary containing $d=95$k atoms, was simulated from EPG~\cite{weigel2015epg} and used in methods that required quantitative mapping. Except the SCQ baseline, other methods used a time-domain linear dimentionality reduction with $s=5$ from~\cite{mcgivney2014svdmrf}. 
Reconstruction performance was evaluated using Mean Average Percentage Error (MAPE) for skull-stripped qmaps, channel-averaged Normalized Root Mean Squared Error (NRMSE) for TSMI, and NRMSE between k-space measurements ($\y$) and predictions ($\y_\text{rec} := \mathcal{A} \x_\text{rec} $), where applicable. 
Experiments ran on an NVIDIA GeForce RTX 4090 GPU. 
\newline
\newline
\textbf{Baseline methods:} 
Our method was evaluated against SVDMRF~\cite{mcgivney2014svdmrf}, MRF-ADMM~\cite{asslander2018admm}, LRTV~\cite{golbabaee2021lrtv}, SCQ~\cite{fang2019supervisedmrf} and MRF-IDDPM~\cite{mayo2024ddpmmrf}. 
\textbf{SVDMRF} uses $\x_\text{rec}:=\mathcal{A}^H \y$ to reconstruct the TSMIs (equivalent to the conditions $\x_c$ in our approach) which are then passed to \DM\ for quantitative mapping. 
\textbf{MRF-ADMM} uses the ADMM algorithm to enforce k-space consistency and Bloch constraints (\DM) during reconstruction. Here we did not employ additional regularisation. \textbf{LRTV} employs k-space consistency and a Total Variation image regularization using convex optimisation.  
\textbf{SCQ} is a CNN-based deep learning method that maps aliasing-contaminated TSMIs from undersampled acquisitions to restored T1/T2 maps. It uses a fully-connected network to reduce the TSMI dimension, then employs two UNets to estimate T1 and T2 maps separately.
We implemented SCQ following \cite{fang2019supervisedmrf} for architecture, and trained the models for 1k epochs using data augmentation (patching, vertical/horizontal flips),  ADAM optimizer, MSE loss, and a learning rate of 0.005--parameters that pefermed best for our data. 
\textbf{MRF-IDDPM} is a conditional DDM model that does not employ physics-driven guidance during sampling. We assess its sampling performance using the same network $\epsilon_\theta$ trained for MRF-DiPh. 
\newline
\newline
\textbf{Ablations:}
\label{subsec:ablations}
The parameters $\lambda$ and $\tau$ were tuned to reach a satisfactory balance between image prior and physics guidance. We found $\lambda=10^{-4}$, and $\gamma \in [0.01,0.1]$ provided competitive performance. To assess the effect of $\xi$, we tested two MRF-DiPh sampling configurations, one used an even mix of deterministic and stochastic noise ($\xi=0.5$, Mode A), whereas a second used purely deterministic noise ($\xi =0$, Mode B), and found no significant impact (Table~\ref{tab:results}). 
Two additional modes were evaluated: Mode C, which combined DDM sampling with k-space consistency but without dictionary matching, and Mode D, which employed an unconditional DDM denoiser trained separately without conditioning images $\x_c$, while keeping other training parameters the same as our base model. 
To improve sampling speed, we inspected the maximum number of CG iterations in $prox_f$, and the number sampling steps $K$. Results of these investigations are in Tables~\ref{tab:results} and~\ref{tab:ablations}. 
\section{Results}
\label{sec:results}
\begin{figure}[t!]
    \centering
    \includegraphics[width=\linewidth,trim={0.3cm 0.25cm 0.25cm 0.25cm},clip]{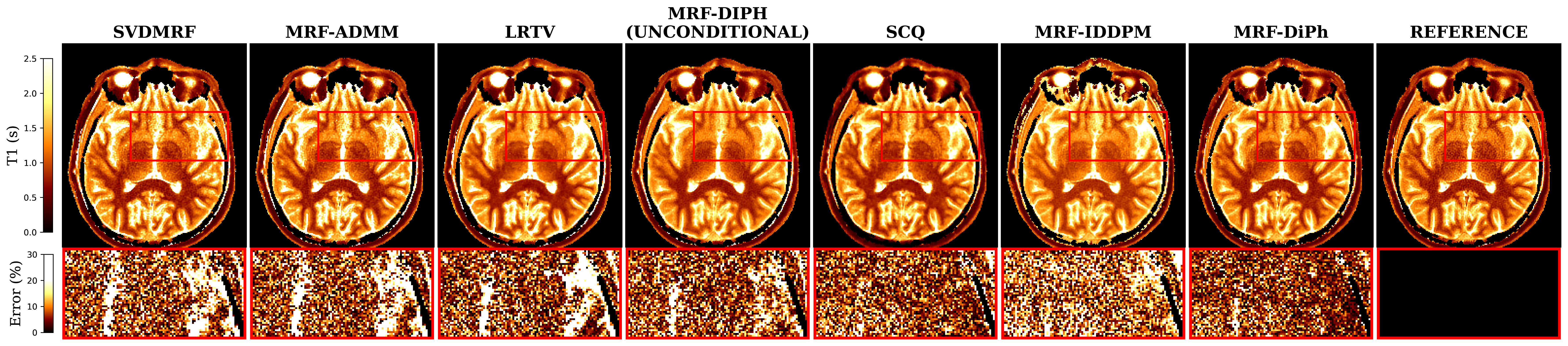} 
    \includegraphics[width=\linewidth,trim={0.25cm 0.25cm 0.25cm 0.7cm},clip]{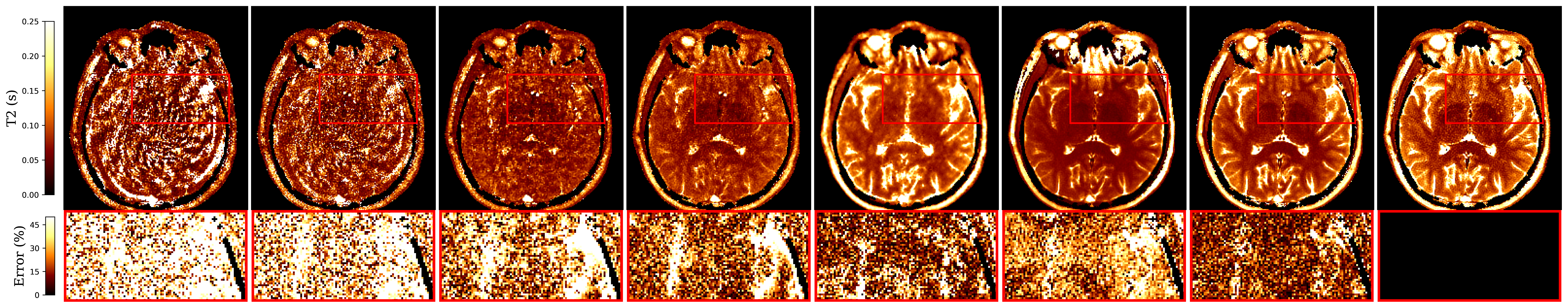} 
    \caption{Reconstructed T1-T2 maps and their zoomed-in absolute percentage error maps (rows), using different methods (columns) for a representative brain slice in test set (electronic zoom recommended). 
    }
    \label{fig:qmaps}
\end{figure}

MRF-DiPh outperforms the tested baselines across all  reconstruction metrics (Table~\ref{tab:results}), improving T1 MAPE by $\sim$2\%, T2 MAPE by $\sim4.2\%$, and TSMI NRMSE by $~\sim8.6\%$. Figure~\ref{fig:qmaps} compares the reconstructed T1 and T2 maps, showing MRF-DiPh with fewer errors and more defined anatomies. 
Table~\ref{tab:results} shows that MRF-DiPh outperforms the purely data-driven MRF-IDDPM not only in quantitative mapping, but also in k-space fidelity by $\sim 13\%$ less errors, leveraging physics-informed guidance during reconstruction.
Enforcing only k-space consistency along with DDM (ablated case C) improved performance over MRF-IDDPM, while adding Bloch consistency (MRF-DiPh base) further enhanced qmap reconstructions.
A trade-off exists with k-space consistency--weak or absent image priors can lead to overfitting, introducing artifacts from noisy undersampled data (see LRTV, ADMM-MRF).

In our experiments, the unconditional model (MRF-DiPh D) improved upon SVDMRF, ADMM-MRF and LRTV methods. However, it underestimated T2 maps, particularly around grey matter and CSF areas (Fig.~\ref{fig:qmaps}), falling short of conditional/supervised learning baselines. The embedded conditioning in our base model
was able to resolve this challenging task. Therefore, exploring the application of unconditional MRF-DiPh in other acceleration schemes -- such as further k-space subsampling, rather than truncating the sequence length -- requires further investigation.

\begin{table}[t!]
    \fontsize{8pt}{8pt}
    \caption{
    Reconstruction metrics (averaged over the test dataset) for TSMI, T1 and T2 maps, as well as the k-space fitting errors for different methods.
    }
    \label{tab:results}
    \centering
    \begin{tabular}{l|cc|cc}
    \hline
         \textbf{\footnotesize{Method}  }& \textbf{  MAPE T1  } & \textbf{  MAPE T2  } & \textbf{ 
 NRMSE TSMI  } & \textbf{  NRMSE K-Space  }  \\ \hline
         \textbf{SVDMRF}   & 20.01 & 144.27 & 57.10 & 99.93   \\
         \textbf{MRF-ADMM} & 20.30 & 68.51 & 30.90 & 18.69 \\
         \textbf{LRTV}     & 19.84 & 39.01 & 37.95 & 11.72  \\
         \textbf{SCQ}      &  8.76 & 22.61 &  -- & -- \\ 
         \textbf{MRF-IDDPM}    & 8.45 & 22.54 & 27.26 & 36.06 \\
         \hline
         \textbf{MRF-DiPh (base)} & 6.75 & 18.40  & 18.65 & 22.82  \\ 
         \hline
         \textbf{MRF-DiPh (A)} & 6.80 & 18.41 & 18.70 & 22.52  \\ 
         \textbf{MRF-DiPh (B)} & 7.15 & 18.63 & 18.64 & 22.40  \\ 
         \textbf{MRF-DiPh (C)} & 7.17 & 18.82 & 18.79 & 22.12  \\ 
         \textbf{MRF-DiPh (D)}  & 11.32 & 29.78 & 25.36 & 19.17 \\ 
         \hline
    \end{tabular}
\end{table}

\textbf{The runtimes} 
MRF-DiPh base model reconstructions took $\sim44$ seconds, slightly longer but comparable to the iterative baselines LRTV and ADMM-MRF. The three major steps, DDM denoising, k-space consistency ($prox_f$) and \DM\ took approximately 7\%, 53\% and 40\% of the total runtime. 
Reconstruction time can be decreased by reducing the number of CG iterations in $prox_f$ (e.g. to 1), or lower sampling steps $K$ (e.g. to 20), allowing a controlled trade-off with accuracy (see Table~\ref{tab:ablations}). Exploring faster/approximate 
\DM\ alternatives~\cite{cauley2015fastgdm,golbabaee2019coverblip} could further enhance efficiency. 
Our model took 55.5 hours to complete 100k training iterations. However, analysis of the performance checkpoints reveals that MRF-DiPh, utilizing physics-based guidance, achieves competitive test accuracy early in training—significantly faster than the purely data-driven MRF-IDDPM model. This advantage could have significant implications for 3D/high-dimensional imaging tasks, where training becomes highly resource-intensive.

\begin{table}[t!]
    \fontsize{8pt}{8pt}
    \caption{MRF-DiPh reconstruction time vs. accuracy for the options: number of sampling steps (K), and maximum CG iterations. Results for the test image in Fig~\ref{fig:qmaps}.}
    \label{tab:ablations}
    \centering
    \footnotesize
    \begin{tabular}{l|c|cccc|ccc}
    \hline
    \multirow{ 2}{*}{\textbf{MRF-DiPh}} & \multirow{ 2}{*}{\textbf{      Base     }} & \multicolumn{4}{ c|}{\textbf{K}}  & \multicolumn{3}{ c}{\textbf{CG}} \\ \cline{3-9}
     &  & \textbf{      5     } & \textbf{      10     } & \textbf{     20     } & \textbf{     50     } & \textbf{     1     } & \textbf{     10     } & \textbf{     20    } \\
    \hline
    \textbf{Runtime (s)} & 44.17 & 8.83 & 15.88 & 30.03 & 72.29  & 30.02  & 61.03  & 91.79 \\ \hline
    \textbf{TSMI NRMSE} &  15.76 & 16.74 & 14.99 & 15.55 & 16.06 & 15.73 & 15.87 & 15.70  \\
    \hline
    \textbf{(T1+T2)/2 MAPE  } & 10.42 & 14.20 & 11.93 & 10.62 & 10.22 & 10.37 & 10.45 & 10.33  \\

    \hline
    \end{tabular}
\end{table}

\begin{table}[t!]
    \fontsize{8pt}{8pt}
    \caption{Average reconstruction errors (MAPE T1 + MAPE T2)/2 over the test set throughout training iterations for MRF-DiPh and baseline MRF-IDDPM.}
    \label{tab:ckp}
    \centering
    \begin{tabular}{l|cccccccccc}
    \hline
    \textbf{Checkpoint} $\mathbf{\times 10^3}$ & 10 & 20 & 30 & 40 & 50 & 60 & 70 & 80& 90& 100 \\
    \hline
    \textbf{MRF-DiPh} & 145.81 & 23.70 & 14.41 & 13.30 & 12.87 & 12.62 & 12.48 & 12.44 & 12.57 & 12.58  \\
    \hline
    \textbf{MRF-IDDPM} & 340.51 & 347.01 & 405.90 & 404.77 & 225.16 & 128.07 & 60.53 & 25.22 & 19.06 & 15.50 \\ 
    \hline
    \end{tabular}
\end{table}

\section{Conclusions}
\label{sec:conclusions}
This work introduced MRF-DiPh, a diffusion-based model for reconstructing MRF data with more accurate tissue parameter estimations and improved measurement fidelity. The proposed method intertwines the sampling steps of the diffusion model with measurement and Bloch consistency regularizations. Our experiments demonstrate MRI-DiPh's potential with robust reconstructions that incorporate physics-driven  guidance.

\begin{credits}
\subsubsection{\ackname} This work was supported by the EPSRC grant EP/X001091/1. 
\end{credits}

%
%
%
\bibliographystyle{splncs04}
\bibliography{references}

\end{document}